\documentclass{appolb}
\usepackage{graphicx,epsfig}
\usepackage[utf8]{inputenc}
\usepackage{amsmath}

\begin{document}
\title{Correspondence of Many-flavor Limit and Kaluza\,--\,Klein Degrees of Freedom in the Description of Compact Stars%
}
\author{Szilvia Karsai$^{a,b}$, Gergely G\'abor Barnaf\"oldi$^{a}$, \\
Emese Forg\'acs-Dajka$^{b}$, P\'eter P\'osfay$^{a,b}$ 
\address{$^a$Wigner Research Centre for Physics of the H.A.S., \\P.O. Box 49, Budapest H-1525, Hungary, \\
$^b$  E\"otv\"os Lor\'and University, Department of Astronomy, \\ P\'azm\'any P\'eter s\'et\'any 1/A, Budapest H-1117, Hungary}
}
\maketitle
\begin{abstract}
We present the correspondence between non-interacting multi-hadron fermion star equation of state in the many-flavor limit and the degrees of freedom of a Kaluza\,--\,Klein compact star. Many flavors can be interpreted in this framework as one extra compacti\-fied spatial dimension with various, more-and-more massive hadron state excitations. The effect of increasing the degrees of freedom was investigated on the equation of state and in connection with the mass-radius relation, $M(R)$. The maximum mass of the star, $M_{\mathrm{max}}$ were also calculated as a function of the maximum number of excited states, $n$ and the size of the compactified extra dimension, $R_{\mathrm{c}}$.
\end{abstract}

%
%
%
%
\section{Introduction}
\label{sec:intro}

Investigation of the phase diagram of hot and dense matter aims to cla\-rify the role of phase transitions between partonic and hadronic states, search for the critical point, and explore the exotic degrees of freedom predicted by nuclear physics and low-energy Quantum Chomodynamics (QCD). Lattice QCD works well in the high-temperature and low-density part of the phase diagram while, high-energy heavy-ion collisions have the focus on the critical point. Testing the cold, superdense, strongly interacting matter in the zero-temperature and high-density limit, compact stars are the best (celestial) laboratories, especially through the observation of astrophysical properties of these extreme objects. 

However, there is no direct observation of the inner structure of a compact star, some physical properties like the measured mass-radius relation, moment of inertia, rotation period, magnetic field and the soon-to-be-available gravitational wave observations support to build and parametrize realistic equations of state (EoS) in the non-perturbative and high-density QCD regime. This is one of the main goal of the Working Group 2 of the "NewCompStar COST Action MP1304", which developed a continously-evolving online database, CompOSE~\cite{composeweb} for neutron star EoS.

The aim of this paper is to present the correspondence between non-interacting multi-hadron fermion star equation of state in the many-flavor limit to the many degrees of freedom of a Kaluza\,--\,Klein compact star. We introduce the description of a compact star in the Kaluza\,--\,Klein world with one extra compactified dimension and show how excited states can be connected to the hadron spectra, especially in the many-flavor limit, assuming more-and-more massive hadron states. 

We present the effect of increasing the degrees of freedom on the equation of state and the connection of this conception to the mass-radius relation, $M(R)$ in compact stars. We present the dependence of the maxi\-mum mass of the star, $M_{\mathrm{max}}$ as a function of the maximum number of excited states, $n$ and the size of the compactified extra dimension, $R_{\mathrm{c}}$.

\section{Degrees of Freedom in the Kaluza\,--\,Klein Theory}
\label{sec:kk-star}

In the Kaluza\,--\,Klein model the gravity and quantum field theory can be unified at energy scale lower than the Planck's scale~\cite{Kaluza:1921,Klein:1926,Antoniadis:1990}. In the simplest case a $3+1_{\mathrm{c}}+1$ dimensional space-time can be introduced, where excited particles can move freely along the extra $x^5$ spatial direction as well. In this manner we 'geometrize' quantum fields, where charges are associated with compactified spatial extra dimensions, induced by the topological structure of the space-time. 

The compactness of the extra dimension generates a periodic boundary condition, which results a Bohr-type quantization condition for the $k_5$ momentum component. Applying a generalized Heisenberg uncertainty formula, the relation induces an uncertainty in the position with the size (volume) of $2 \pi R_{\mathrm{c}}$. An interesting feature of this space-time structure, that motion into the 5\textsuperscript{th} dimension generates an extra mass term by $k_5$, what appears as 'excited mass', $m$ in the standard $3+1$-dimensional space-time, 
\begin{equation}
k_5 = { n \,\,\hbar c}/{R_{\mathrm{c}}} \,\,\,\,\,\,\,\ 
\longrightarrow \,\,\,\,\,\,\,\, 
m=\sqrt{\widehat{m}^2+\left({n \,\,\hbar c}/{R_{\mathrm{c}}}\right)^2 }, 
\label{quanta}
\end{equation}
where $\widehat{m}$ is the particle mass in the 5-dimensional description, and $n$ is the excitation number. Considering a compactified radius $R_{\mathrm{c}} \sim 10^{-13}$ cm this 'extra mass' gap is $ \sim 100$ MeV, which is an available value in hadron spectroscopy by the TeV energy accelerators, such as the Large Hadron Collider~\cite{Arkhipov:2004} or even in superdense compact stars as in Ref.~\cite{Barnafoldi:2007} and references therein. 
In this framework the extra compactified 5\textsuperscript{th} dimension represents the hypercharge or the similar quantum number (strangeness, charm, or bottomness) connected to even several number of flavors.

The masses of the Kaluza\,--\,Klein excited states (or degrees of freedom) can be calculated by the formula~\eqref{quanta}. In Table~\ref{tab:kk-masses} we listed these masses in MeV units up to excitation number $n\leq 10$ and compactified radius values for $R_{\mathrm{c}}=$0.01, 0.1, 0.33, 0.5, 1.0, and 10 fm with $m_{n=0}=940$ MeV. One can observe, higher $n$ results more massive baryons, but larger $R_{\mathrm{c}}$ turns excited-state structure more dense to each other like in case of many flavor.  
%
%
\begin{table}[!h]
\centering
\label{tab:kk-masses}
\begin{tabular}{lccccccccc}
\hline
Exct.  &  & &  $R_{\mathrm{c}}$ [fm] &  \\
$n$ & 0.01 & 0.1 & 0.33 & 0.5 & 1.0 & 10   \\
\hline
\hline
0 & 940.0 & 940.0 & 940.0 & 940.0 & 940.0 & 940.0 \\
1 & 19755.0 & 2185.7 & 1114.0 & 1019.5 & 960.5  & 940.2  \\
2 & 39476.4 & 4056.9 & 1521.1 & 1227.4 & 1019.5 & 940.8  \\
3 & 59205.3 & 5993.9 & 2025.2 & 1511.7 & 1110.8 & 941.8  \\
4 & 78936.0 & 7948.8 & 2569.9 & 1837.3 & 1227.4  & 943.3  \\
5 & 98667.5 & 9911.0 & 3134.1 & 2185.7 & 1362.7  & 945.2  \\
10 & 197328.0 & 19755.0 & 6053.0 & 4056.9 & 2185.7  & 960.5  \\
\hline
\end{tabular}
\caption{Masses of Kaluza\,--\,Klein degrees of freedom in MeV units, $m(n,R_{\mathrm{c}})$.}
\end{table}
%

\section{Results for Compact Stars: EoS with Many-flavor }
\label{sec:kk-eos}

\begin{figure}[!h]
\begin{center}
\resizebox{0.29\textwidth}{!}{\includegraphics{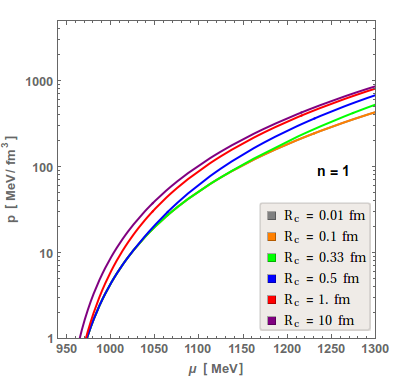}}
\resizebox{0.29\textwidth}{!}{\includegraphics{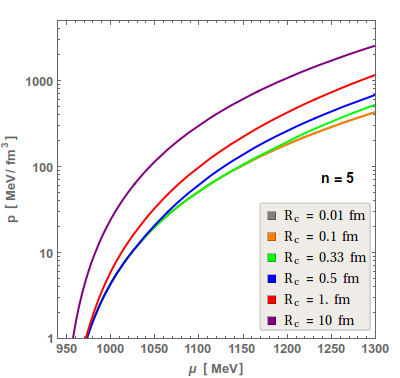}}
\resizebox{0.29\textwidth}{!}{\includegraphics{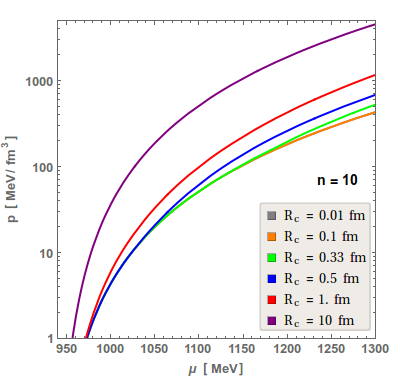}}
\resizebox{0.29\textwidth}{!}{\includegraphics{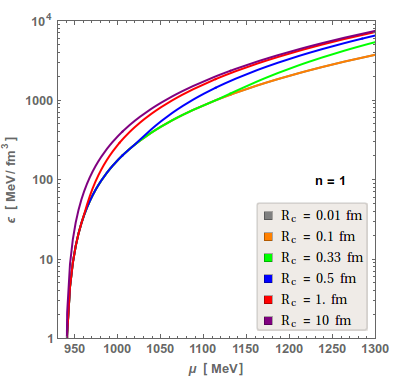}}
\resizebox{0.29\textwidth}{!}{\includegraphics{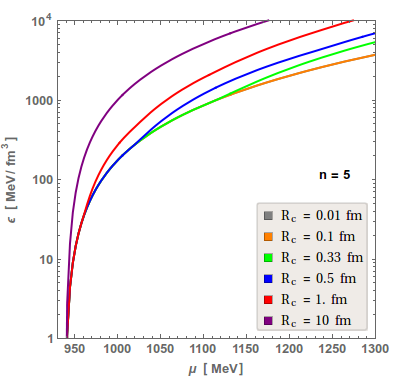}}
\resizebox{0.29\textwidth}{!}{\includegraphics{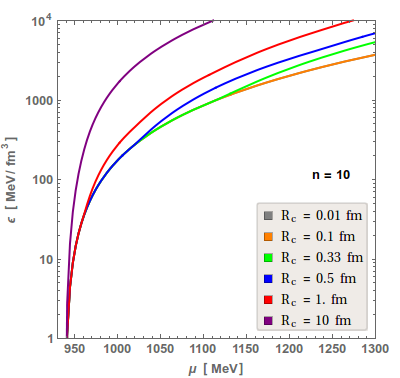}}
\resizebox{0.29\textwidth}{!}{\includegraphics{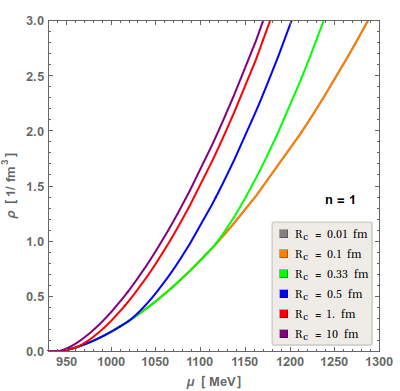}}
\resizebox{0.29\textwidth}{!}{\includegraphics{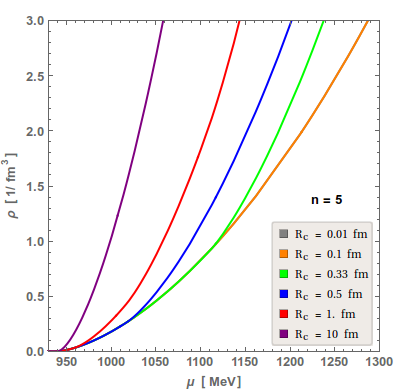}}
\resizebox{0.29\textwidth}{!}{\includegraphics{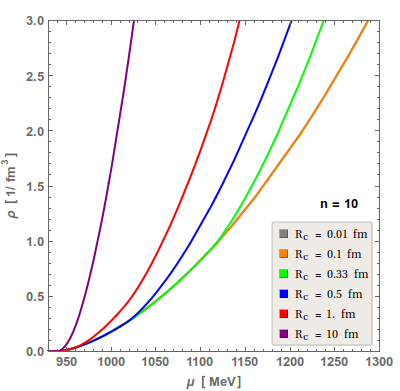}}
\resizebox{0.30\textwidth}{!}{\includegraphics{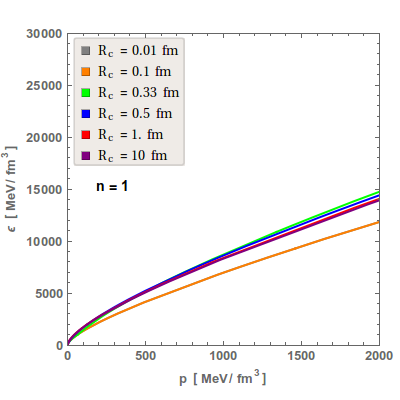}}
\resizebox{0.30\textwidth}{!}{\includegraphics{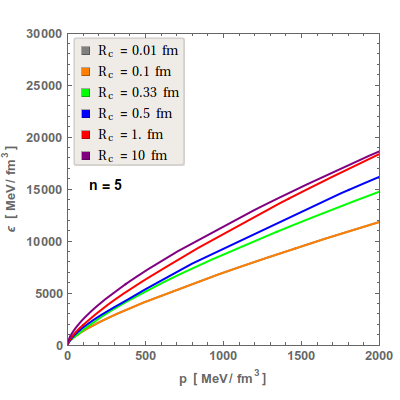}}
\resizebox{0.30\textwidth}{!}{\includegraphics{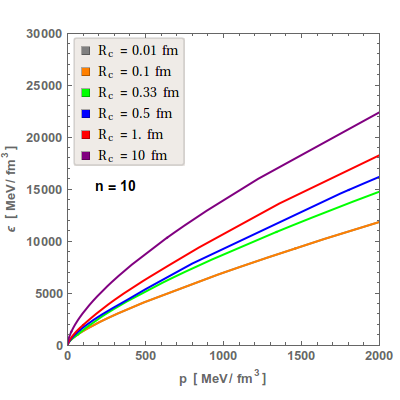}}
\resizebox{0.305\textwidth}{!}{\includegraphics{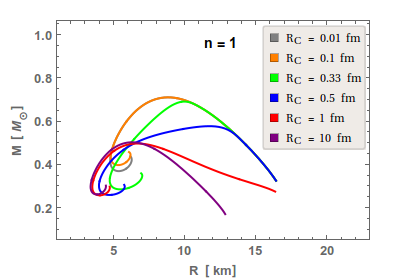}}
\resizebox{0.305\textwidth}{!}{\includegraphics{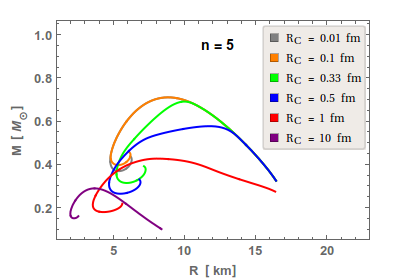}}
\resizebox{0.305\textwidth}{!}{\includegraphics{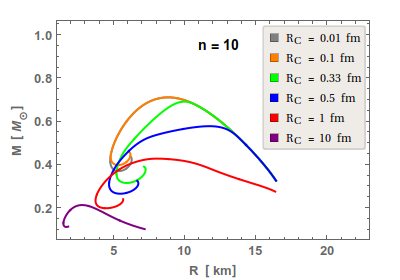}}
\end{center}
\vspace*{-0.5truecm}
\caption{The EoS dependence on the Kaluza\,--\,Klein excites states as a function of $R_{\mathrm{c}} =$ 0.01, 0.1, 0.33, 0.5, 1.0, 10 fm and excitation number $n=1$, 5, and 10 values.}  
\label{fig:eos-mr}
\end{figure}

Based on the above model on the degrees of freedom, we present our results for the EoS in the case of non-interacting fermion gas in $3+1_{\mathrm{c}}+1$ dimensional space-time with excitations, $n$ and $R_{\mathrm{c}}$ values as highlighted above. On Fig.~\ref{fig:eos-mr} the calculated $p(\mu)$, $\varepsilon(\mu)$, $\rho(\mu)$, and $\varepsilon(p)$ curves are plotted from the top to the bottom panels respectively. The above thermodynamical variables were derived for any states $i$, from the generalized thermodynamical potential, $\Omega (m_i,\mu_i)$ as presented in Refs.~\cite{Barnafoldi:2007,GribovProc:2015}. 

In general, one reads from the graphs that increasing the $R_{\mathrm{c}}$ makes softer the EoS, and {\sl vice versa}, $R_{\mathrm{c}} \to 0$ results stiffer one. Moreover, for the low-$\varepsilon$/low-$p$ sector, increasing the number of possible excitations, $n$, does not make any sense with the EoS with the smallest $R_{\mathrm{c}}$ values, on the other hand, at larger $R_{\mathrm{c}}$ values, the opening of the large number of excited states results more softer EoS. This is due to the well-populated d.o.f. structure at the low-mass baryon sector. 

The bottom line of Fig.~\ref{fig:eos-mr} presents the mass-radius relation, $M(R)$ of compact star models calculated from Tolman\,--\,Oppenheimer\,--\,Volkov equation in static, spherical symmetric, 5-dimensional spacetime applying several many-component, non-interacting fermion EoS-s for given $n$ and $R_{\mathrm{c}}$ values.

\begin{figure}[!h]
\begin{center}
\resizebox{0.5\textwidth}{!}{\includegraphics{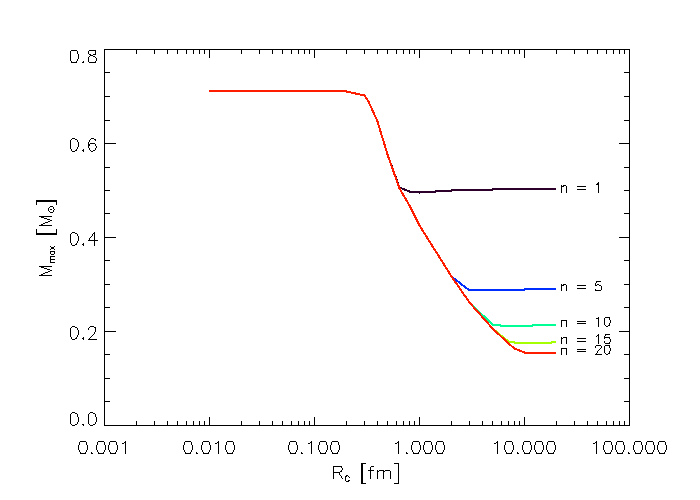}}
\end{center}
\vspace*{-0.5truecm}
\caption{The maximum mass of a compact star, $ M_{\mathrm{max}} $ as a function of $R_{\mathrm{c}}$ and $n$.}  
\label{fig:Mmax-Rc-n-dep}
\end{figure}

The $M(R)$ diagram clearly presents that decreasing the $R_{\mathrm{c}}$, the maxi\-mum mass, $M_{\mathrm{max}}$ of the star is getting larger, since the EoS of the star becomes stiffer. As getting $R_{\mathrm{c}}\to 0$ the $M_{\mathrm{max}}$ increases and saturates to a maximum $0.7 M_{\odot}$, independently of the $n$. As increasing the $R_{\mathrm{c}}$ (softens the EoS), $M_{\mathrm{max}}$ also saturates, but results smaller-and-smaller $M_{\mathrm{max}}$ depending on the possible degrees of freedom, $n$, as it is summarized on Fig.~\ref{fig:Mmax-Rc-n-dep}.  
\vspace*{-0.3truecm}

\vspace*{-0.3truecm}
\section{Summary}
\label{sec:summary}

Correspondence between non-interacting multi-hadron fermion star EoS in the infinite flavor limit to the degrees of freedom of a Kaluza\,--\,Klein compact star has been presented. We found that decreasing $R_{\mathrm{c}}$ results stiffer EoS with saturated $M_{\mathrm{max}} \to 0.7 M_{\odot} $, while larger $R_{\mathrm{c}}$ generates saturations at smaller $M_{\mathrm{max}}$ depending on the number of degrees of freedom (flavors).

\section*{Acknowledgments}

This work was supported by NewCompStar COST action MP1304, Hungarian OTKA grants NK106119, K104260, K104292, TET 12 CN-1-2012-0016, and the J\'anos Bolyai Research Scholarship of the H.A.S. (GGB).



\vspace*{-0.3truecm}
\begin{thebibliography}{99}  

\bibitem{composeweb}
CompOSE Website {\tt http://compose.obspm.fr/}

\bibitem{Kaluza:1921}
Th. Kaluza, 
{\it Sitzungsber Preuss. Akad. Wiss. Phys. Math Kl.} {\bf LIV}, 966 (1921).

\bibitem{Klein:1926}
O. Klein, 
{\it Z. Phys.} {\bf 37}, 875 (1926).

\bibitem{Antoniadis:1990}
I. Antoniadis, 
{\it Phys.Let.} {\bf B} 246 3,4 (1990). 

\bibitem{Arkhipov:2004}
A. Arkhipov,
{\tt arXiv:astro-ph/0309327v5}. 

\bibitem{Barnafoldi:2007}
G.G. Barnaf\"oldi, B. Luk\'acs, and P. L\'evai, 
{\it AN} {\bf 328}, 809 (2007).

\bibitem{GribovProc:2015}
G.G. Barnaf\"oldi, Sz. Karsai, B. Luk\'acs, P. P\'osfay,
{\tt arXiv:1509.07354v1}.

\end{thebibliography}
\end{document}